%% file: main.tex
\documentclass[12pt, a4paper, oneside]{article}
\usepackage{paper-template}
\usepackage{tikz}
\usepackage{enumitem}
\usetikzlibrary{arrows.meta, positioning, fit, backgrounds, calc}

\title{%
  \vspace{-6mm}
  {\large\bfseries Artificial Intelligence in Ship Finance: Applications, Opportunities, and a Case Study in AI-Augmented Loan Origination}%
  \vspace{2mm}
}

\author{%
  \normalsize
  \textbf{Lasse Dierich}$^{1, 2, 3}$, \textbf{Orestis Schinas}$^{1, 2, 4}$\\[2pt]
  \small $^{1}$ShipFinance.ai, Munich, Germany\\
  \small $^{2}$HHX.blue GmbH, Kiel, Germany\\
  \small $^{3}$Technical University of Munich, Munich, Germany\\
  \small $^{4}$University of the Aegean, Chios, Greece\\
  \small \texttt{lasse.dierich@tum.de, oschinas@aegean.gr}
}

\date{}

\begin{document}

\maketitle
\thispagestyle{fancy}

\begin{abstract}
\noindent
Ship finance is a data-intensive and document-heavy segment of asset-based lending, requiring the integration of financial, technical, contractual, and regulatory information from heterogeneous and largely unstructured sources. Increasing environmental regulation and ESG reporting requirements are adding further complexity to underwriting and loan-origination processes. Recent advances in artificial intelligence (AI), particularly large language models (LLMs), create new opportunities for processing and analysing such information.
This paper reviews potential applications of AI in ship finance, with a particular focus on LLM-based systems for document comprehension, information extraction, and workflow automation. We present \textit{ShipFinance.ai}, a modular agentic architecture to support loan application workflows in ship finance. The proposed system combines an LLM-based extraction module, financial analysis components, external maritime data services, and a controlled document-generation module with a chatbot interface to support the preparation of standardized financing applications. The paper discusses the key challenges for using such models in production. We argue that AI-assisted systems can support maritime finance professionals in managing increasingly complex information and reporting requirements.

\end{abstract}

\section{Introduction}
The financing of ships remains a specialized, cyclical, and relationship-driven segment of corporate credit. Its core product is still the senior secured mortgage loan, normally advanced against the asset (the vessel - ship), the employment contract, and the sponsor's balance sheet, as well as personal credibility and reputation. Although shipping is a global industry, the market is substantially fragmented. European, Japanese, Greek and regional banks coexist with Chinese, Japanese and Korean sale-and-leaseback lessors, export-credit agencies, private-credit funds, leasing houses, public equity and Nordic/high-yield bond investors. In 2024, shipping bonds reportedly doubled from US\$7.4~billion to US\$14.5~billion \citep{Petrofin2025,HellenicShippingNews2025}. The older German \emph{Kommanditgesellschaft} (KG) scheme, once a major source of retail equity for single-ship companies in Germany and in similar legal forms in Scandinavia, is now mainly relevant as a legacy structure after the post-2008 collapse of retail KG origination, but it remains useful for understanding why ship finance often separates asset ownership, tax treatment and commercial control \citep{SchinasGrauJohns2015,VDRShipFinance}.

The scale is material even for readers outside shipping. \citet{Petrofin2025} estimates that, at end-2024, the top 40 shipping banks held about US\$289.65~billion of shipping exposure~\citep{Petrofin2025}. The total bank lending, including smaller and local banks, was around US\$400~billion, and the broader ship-finance stock, including leasing, export finance and alternative lenders, was about US\$625~billion \citep{Petrofin2025}. Considering the value of the fleet in the water close to US\$1250~billion, a leverage of 50\% is revealed, which seems reasonable given average market terms. On a flow basis, recent market commentary puts new annual exposure in the high tens of billions of dollars, around US\$88~billion in one estimate \citep{TradeWinds2025}. Individual facilities are therefore not small corporate loans: a single-vessel loan may be only US\$5-20~million for small second-hand tonnage, but US\$30-100~million is common for modern bulkers, tankers or container vessels, while multi-vessel, LNG, cruise or newbuilding facilities can exceed US\$250~million \citep{SewardKisselShipFinanceBasics,HellenicShippingNews2025}. Large tickets, usually over US\$50-70~million, are syndicated, while capital needs over US\$150-170~million may be addressed via the public offering of bonds or shares.

The scale of capital requirements explains why ship finance is both data-intensive and relationship-driven. Credit committees do not evaluate only a borrower, but a moving income-producing asset exposed to freight cycles, fuel prices, charterparty terms, technical performance, class status, flag, insurance, residual value, and enforcement risk across jurisdictions. The loan application can therefore be strengthened by a structured evidence package: audited accounts, charter backlog, vessel valuations, employment scenarios, loan-to-value stress tests, fuel-consumption records, EU Monitoring, Reporting and Verification (EU MRV) and IMO Data Collection System (IMO DCS) submissions, Carbon Intensity Indicator (CII) and Energy Efficiency Existing Ship Index (EEXI) status, retrofit capital expenditure (capex), EU Emissions Trading System (EU ETS) cost sensitivities, sanctions screening, and Environmental, Social, and Governance (ESG) assurance. The better the data quality, the less the lender must rely on conservative assumptions, and the easier it becomes to justify leverage, tenor, and covenant headroom \citep{Schinas2026SustainableFinance,PoseidonPrinciples2024}.

Sustainable-finance rules increase the complexity further by linking vessel emissions to access to capital. The EU Taxonomy, SFDR disclosures, the maritime EU ETS and voluntary frameworks such as the Poseidon Principles require lenders to collect, verify and disclose emissions and alignment data. This turns carbon performance into a credit variable rather than a public-relations issue \citep{EuropeanParliamentCouncil2019,EuropeanParliamentCouncil2020,EuropeanParliamentCouncil2023,EuropeanCommissionETSShipping}. Recent work on ship finance and green-ship funding shows that environmental regulation changes both investment appraisal and financing design, because green technologies often require higher upfront capital while the commercial benefit is uncertain \citep{SchinasRossRossol2018,SchinasMetzger2019,SchinasBergmann2021}. Consequently, rejection, re-pricing or withdrawal of loan applications is costly: the borrower may lose a charter window, yard slot or second-hand acquisition, while the lender spends scarce underwriting capacity on incomplete cases. In a market where charter opportunities can arise suddenly, for example when supply-chain disruption creates immediate demand for tonnage, a faster and better-substantiated application has direct economic value. Improving the probability of credit approval is therefore not merely administrative efficiency but a source of competitive advantage.

Over the past decade, artificial intelligence (AI) systems have increasingly been adopted to automate document-heavy and decision-support workflows in finance \citep{AI_in_financial_services_survey}. Recent advances in Large Language Models (LLMs) have substantially expanded the ability of AI systems to process unstructured textual information, including contracts, financial reports, and regulatory filings \citep{AI_integration_in_financial_services_overview, LLMs_in_finance}. Specialised LLMs like BloombergGPT \citep{BloombergGPT} and FinGPT \citep{FinGPT} have been trained on financial documents for an improved understanding of the language used in them. These developments are particularly relevant for ship finance, where underwriting processes require the integration of heterogeneous document sources with many specialist terms.
Adjacent works focus on credit rating criteria for ship finance~\citep{integrated_credit_rating_and_loan_quality_model}, or apply machine learning methods like Graph Neural Networks (GNNs) and Support Vector Machines (SVMs) to predict loan default~\citep{loan_default_prediction_topological_data_analysis_and_machine_learning}. To the best of our knowledge, no previous work addresses the application of the latest developments in AI, specifically LLMs, to ship finance. We aim to bridge this gap by reviewing possible applications and opportunities for AI in ship finance and present a case study on AI-augmented loan origination in ship finance.

\section{AI: a primer for the maritime finance professional}

\subsection{Defining the terms}
AI can best be understood as a type of technology in which computers think or act like humans, allowing them to perform tasks that would traditionally have required human intelligence to complete~\citep{russell_norvig_modern_ai}. Machine Learning (ML) is a broad subfield of AI, focusing on the learning of patterns in data, allowing many industry-relevant applications such as classification tasks or anomaly detection~\citep{russell_norvig_modern_ai}. Natural language processing (NLP) is a field of AI concerned with enabling computers to perform tasks involving human language, including language understanding and generation~\citep{nlp_jurafsky}. For example, Large Language Models (LLMs) are transformer-based foundation models trained on large-scale text corpora to perform a wide range of natural language tasks~\citep{bommasani2021foundation,vaswani2017attention}. LLMs gained widespread public attention following the release of chatbot-based interfaces for Generative Pre-trained Transformer (GPT) models, particularly OpenAI's ChatGPT~\citep{ouyang2022training}.

Recent advances have extended LLMs beyond single-turn question answering into \emph{agentic} frameworks \citep{agentic_ai_comprehensive_overview}, incorporating retrieval-augmented generation \citep{lewis2020rag} to ground responses in external knowledge bases, and equipping models with tools such as database lookups, calculation engines, or external API calls. Agentic frameworks reason in multiple steps, select the appropriate tools for sub-tasks, inspect intermediate outputs, and iterate until a goal is achieved \citep{agentic_ai_comprehensive_overview}.

\subsection{AI in the financial sector}
AI has been widely adopted for many tasks in the financial sector.  Customer chatbots and process automation help banks to save costs and time 
\citep{increasing_customer_service_efficiency_through_artificial_intelligence_chatbot, toward_a_chatbot_for_financial_sustainability},  AI in anti-money-laundering monitoring can help banks to react to increasing regulatory pressure \citep{AI_for_AML_in_banking, deploying_artificial_intelligence_for_anti_money_laundering_and_asset_recovery_the_dawn_of_a_new_era, the_application_of_artificial_intelligence_in_financial_compliance_management}. AI for credit scoring can help to ensure a fair and objective assessment \citep{fairness_in_credit_scoring}, and applications in algorithmic trading allow reacting to changes in the market much faster than humans are capable of \citep{algorithmic_trading_and_financial_forecasting_using_advanced_artificial_intelligence_methodologies, comprehensive_analysis_of_AI_applications_in_financial_markets}. Fast-growing areas of AI in the financial sector include AI for various underwriting tasks \citep{AI_for_underwriting_tasks} and generative AI copilots for analysts \citep{generative_AI_in_finance_overview, agentic_ai_comprehensive_overview}. We emphasise that the enhancement of systems with AI only creates real value if the systems can be audited \citep{embedding_governance_into_LLM_workflow_architectures_for_enterprise_wide_automation_IBM, fintech_AI_overview}.

\subsection{From data to decision: Where AI creates value in ship finance}
Decisions in ship finance are typically based on a variety of different documents. The borrower might provide loan covenants, charter agreements, or P\&L statements, whereas databases like Clarksons, EMSA THETIS, or Marine Traffic provide access to public data. The documents are scattered across different storage spaces, the company's internal document management systems, or APIs. Only some of the data is available in structured formats, like standardized spreadsheets, or JSON format retrieved from an API call. Most of the documents have no standardized structure and combine text, tables, and figures in a single file. The lack of standardization complicates the usage of the data for workflows typically used in ship finance, like spreadsheet-based cashflow models. Typically, a manual review of each document by a human is needed before the information can be used. We see the cost and time involved with human review as a significant limitation for how well available data sources are exploited in decision-making in ship finance.

LLMs are naturally suited for document comprehension and are actively used in various contexts for summarising text documents and answering questions about them \citep{LLMs_for_text_summarization}. They have successfully been used to extract information from unstructured documents like financial reports consisting of text and tables \citep{extracting_financial_data_from_unstructured_sources_leveraging_LLMs, DocFinQA_a_long_context_financial_reasoning_dataset} and can handle various document layouts \citep{layoutllm_document_understanding, huang2022layoutlmv3}. The integration of LLMs into automated workflows \citep{react_paper} has proven successful and holds great potential to bridge the gap between the availability of unstructured data and data usage common in ship finance.
It should be noted that due to the probabilistic nature of LLMs, even well-designed systems cannot guarantee perfect extraction accuracy, making human oversight a necessary component. Rigorous testing, also covering special cases like vague quantifiers (\textit{at least}, \textit{approximately}), is required before a production deployment of LLMs in automated workflows.

\section{ShipFinance.ai: a case study in AI-augmented loan applications}
\label{sec:shipfinanceai}

Shipping debt financing remains one of the most fragmented and opaque segments of asset finance. Vessel owners and brokers spend weeks pitching identical projects to multiple lenders in different formats. Lenders are overwhelmed with unstandardized teasers and lack efficient tools to triage or source deals. ESG and credit scrutiny are inconsistently applied. Asset liquidity and remarketability are rarely assessed upfront. In this case study, we investigate how an AI-augmented application process can help to address the challenges of the loan origination process. Industry consultation suggests that comprehensive loan applications currently require 4-8 weeks to prepare, with underwriting consuming an additional 6-12 weeks. The preparation phase involves substantial manual effort: document collection, financial data extraction, cash flow modeling, and narrative drafting. We estimate that AI-assisted document extraction and standardized application generation could reduce preparation time by 30-40\%, based on the typical time allocation across these steps. This primarily benefits smaller operators with fewer in-house analysts. The time savings are most significant when charter agreements and financial statements are available in digital formats. However, these estimates require validation through pilot deployments.

\subsection{System architecture}

We present an agentic architecture, \textit{ShipFinance.ai}, for AI-augmented loan applications in ship finance, shown in Figure~\ref{fig:architecture}. The architecture features a modular design to be auditable and transparent \citep{towards_self_regulating_ai}. The agentic approach is particularly suited for the complex task of preparing a loan application, as typically no single document or query captures the full picture required for a credit decision. Instead, a system must orchestrate extraction, computation, and synthesis across heterogeneous sources in a sequence that may not be fully specified in advance. \citet{react_paper} demonstrated with the ReAct architecture that combining LLM-based reasoning with external actions, such as database accesses and tool calls, can improve decision-making performance while producing interpretable reasoning traces. Access to external databases can reduce hallucinations by providing grounded information that the LLM can incorporate into its reasoning and response generation processes. Furthermore, providing LLMs with access to computational modules, such as functions implementing predefined financial formulas, enables the separation of semantic reasoning from numerical computation and can improve the traceability of the overall decision process~\citep{react_paper}. This modularity can be extended further by incorporating multiple specialized LLM-based modules into the architecture, each responsible for a specific task.

ShipFinance.ai consists of the following components:
\begin{enumerate}
  \item \textbf{Chatbot Interface}: A chatbot-based interface guides the user through the application process. The interface requests information, such as the IMO numbers of the vessels to be financed, and enables the upload of supporting documents, including existing charter agreements.
  \item \textbf{Value Extraction Module}: The value extraction module is an LLM-based component that extracts the information required by the analysis modules, such as charter rates, from the provided documents based on predefined extraction schemes. These schemes may include textual descriptions or example-based specifications of the target values. The module extracts \textit{in-context}, i.e., it associates values with their semantic qualifiers, such as the charter period to which a charter rate applies and might interpret vague or approximate formulations (\textit{at least}, \textit{approximately}) in a context-dependent manner, similar to human analysts. The extracted values are stored in a structured representation to enable downstream processing, while their citations (i.e., coordinates) within the source documents are preserved as metadata to ensure traceability and auditability.
  \item \textbf{External Data Services/Providers}: Known information is used to retrieve additional data from external data providers. For example, the IMO number can be used to look up technical specifications of the ships, like vessel deadweight, on IHS Markit. If documents are retrieved, they are processed by the value extraction module to make their contents available to downstream components.
  \item \textbf{Analysis Modules}: A collection of analysis modules evaluates the extracted and retrieved data and computes the financial indicators required for the application process. These include
  \begin{enumerate}[label=(\roman*)]
    \item A cash flow module that models vessel cash flows in continuous time while accounting for time-dependent regulatory and operational variables, such as EU ETS costs and CII metrics.
    \item An energy efficiency \& emissions module that integrates data from multiple sources: EU MRV submissions, vessel technical specifications, and operational data from AIS tracking services. CII ratings are calculated according to IMO MEPC.328(76) methodology, incorporating actual fuel consumption, distance traveled, and vessel deadweight. EU ETS costs are projected based on extracted voyage patterns, current EU allowance (EUA) prices from market data feeds, and regulatory phase-in schedules (2024: 40\%, 2025: 70\%, 2026: 100\% of emissions). The module flags vessels at risk of CII downgrades (D or E ratings) and quantifies the financial impact of potential operational constraints.
    \item A revenue module that computes the expected revenues from market estimates, and the information provided on charter agreements.
    \item An asset module provides an asset valuation and risk assessment.
  \end{enumerate}
  To ensure auditability, each value is returned with metadata, tracking the citations of the values used to compute it.
  
  \item \textbf{Application Composer}: The application composer aggregates the extracted information and the outputs of the analysis modules into a standardized financing application. To ensure auditability, all numeric values used by the composer are cited with their source(s) (e.g., "Charter\_Agreement.pdf, page 5, line 10", "Clarksons API, retrieved 2026-05-20"). Cases of conflicting values are surfaced rather than silently resolved.
\end{enumerate}
Preliminary testing demonstrated the technical feasibility of using LLMs to extract key financial metrics from term sheets and charter agreements, and the integration with the cash flow model. Further work is required to benchmark extraction accuracy and evaluate robustness for different document types.

\input{fig_architecture}
\subsection{Possible Extensions}

The proposed architecture could be extended further by incorporating an LLM-as-a-judge \citep{self_refine_iterative} module that evaluates generated applications against predefined qualitative criteria. Such criteria could be derived from historical decision patterns or expert-defined underwriting guidelines and would thereby approximate aspects of the qualitative assessment typically performed by analysts on the lender side. By establishing a feedback loop between the evaluation component and the application composer, the system could iteratively refine generated applications with respect to predefined acceptance criteria.

The architecture can furthermore be generalized beyond loan application generation by defining additional extraction schemes, analysis modules, and document composers for other document classes and reporting workflows. One potential application area is ESG and regulatory reporting. Financial institutions aligned with frameworks such as the Poseidon Principles are required to continuously monitor metrics including CII ratings, EU ETS exposure, and fuel-consumption records across financed vessel portfolios \citep{poseidon2019principles}. A monitoring system could ingest MRV and DCS submissions, extract the relevant metrics using an LLM-based extraction module, evaluate them against applicable decarbonisation trajectories, and identify vessels that deviate from targeted alignment pathways. Such processes currently require substantial manual effort, particularly for large multi-vessel portfolios.

A comparable architecture could also support Know-Your-Customer (KYC) and Anti-Money Laundering (AML) monitoring workflows. In such a setting, an agentic monitoring system could continuously evaluate news feeds, sanctions lists, and company registries and cross-reference them against a lender’s portfolio. This could enable compliance officers to identify material changes and potential risk indicators more rapidly than traditional periodic manual review processes.

\section{Challenges, risks, and ethical considerations}

\subsection{Model reliability and lender acceptance}
The suggested architecture is designed to show interpretable value and decision traces. However, the probabilistic nature of LLMs remains a challenge as it cannot be guaranteed that the LLM will always interpret the documents correctly or generate the application texts as expected. While the reliability can be improved by introducing sanity checks for the extracted values and LLM-based evaluation of the generated application texts, the suggested architecture should only be understood as a tool assisting the ship finance professional rather than a replacement for it. It is an open question how well lenders will accept the generated standardized application documents, or whether they would simply be flagged as suspicious because their standardization makes them appear as if they were generated by a machine. We see transparency of the model architecture for lenders as a key requirement for their acceptance. Transparency for lenders might be achieved by sharing summaries of decision traces or even generating supplementary spreadsheets, allowing them to easily verify the calculations used to generate the application. Even without widespread lender acceptance, the architecture can still help borrowers to save time and effort in drafting an application.

\subsection{Cybersecurity and data privacy}

Loan application documents contain some of the most commercially sensitive information that a shipping company produces. Centralising the processing of such material in a cloud-based AI system creates an attractive target for adversarial actors and introduces data-residency questions under GDPR and equivalent regimes. A breach that exposes a borrower's full financial position to a competitor or a hostile counterparty could cause direct commercial harm and expose the operator of the system to significant legal liability. Mitigating these risks requires careful architectural decisions: end-to-end encryption of document uploads and processed outputs, strict access controls with full logging, clear contractual data-processing agreements between the system operator and the underlying model providers, and, where possible, on-premises or private-cloud deployment for the most sensitive processing steps. 

\subsection{Regulatory and ethical considerations}

The EU AI Act provides a comprehensive legal framework for risk classification of AI systems and imposes proportional restrictions \citep{EuropeanParliamentCouncil2024AIAct}. The classification of ShipFinance.ai depends on its role in the lending process. If the system is used by borrowers or advisers to prepare, structure, and evidence a loan application, it is primarily a decision-support and document-generation tool and might be considered of limited risk. In that configuration, it should not make or recommend a binding credit decision, determine loan pricing, or rank borrowers automatically. If, however, a similar architecture were deployed by a regulated lender to evaluate creditworthiness, prioritise applications, or support approval decisions, it most likely qualifies as a high-risk system. The following articles of \cite{EuropeanParliamentCouncil2024AIAct} apply to high-risk systems and should be seen as good practice for the design of AI systems in general:

\begin{description}
    \item\textbf{Article 9 (Risk Management System)}: Continuous identification and mitigation of risks throughout the system's lifecycle, including testing for extraction errors, hallucinations, and bias in training data or algorithmic outputs.
    \item\textbf{Article 10 (Data and Data Governance)}: Documentation of training data sources, data quality verification procedures, and examination of biases. For ship finance, this includes ensuring that historical lending patterns embedded in training data do not systematically disadvantage certain vessel types, flags, or operator demographics.
    \item\textbf{Article 11 (Technical Documentation)}: Comprehensive documentation including architectural design, testing results, performance metrics, known limitations, and human oversight protocols. This documentation must be maintained and updated throughout the system's operational life.
    \item\textbf{Article 13 (Transparency and Information to Users)}: Clear disclosure to borrowers and lenders that AI systems are involved in application preparation or evaluation, including limitations of automated processing and the extent of human oversight in the decision-making process.
    \item\textbf{Article 14 (Human Oversight)}: Implementation of human oversight measures that enable human operators to understand system outputs, monitor performance, and intervene when necessary. For credit applications, this requires qualified personnel who can override or adjust AI-generated content.
\end{description}

Additionally, the integration with financial institutions regulated by authorities like the European Banking Authority or German Federal Financial Supervisory Authority (BaFin) brings the system within the scope of their guidance on loan origination and monitoring \citep{EBA2020LoanOrigination}, or ICT risks in AI use \citep{BaFin2025AICTRisk}. These frameworks emphasize model validation, stress testing, and governance structures that may be more stringent than the AI Act's minimum requirements.

The main ethical risks are not limited to legal compliance. First, automation may create misplaced confidence in apparently well-structured outputs. A tool for preparing polished-looking loan applications may obscure weak data, uncertain assumptions, or missing evidence, and might even be abused intentionally to create a false impression of creditworthiness. Second, bias may enter indirectly through market data, historical lending patterns, sanctions-screening heuristics, or lender-specific acceptance criteria. Third, borrowers on which more data is available may benefit disproportionately, while smaller owners may be disadvantaged if standardized AI-assisted applications become an implicit market norm. These risks can be mitigated through explicit design constraints. The application composer should distinguish verified facts, model-derived estimates, user-provided assumptions, and narrative interpretation. Material uncertainties, missing documents, and conflicting data should be surfaced rather than smoothed over. 

\section{Conclusion}
The financing of ships increasingly requires the integration of heterogeneous information from financial reports, contracts, regulatory filings, and maritime data services. Traditional approaches relying on manual document review create bottlenecks that are particularly burdensome for smaller operators and in fast-moving charter markets where timing can determine whether a financing opportunity succeeds or fails. This paper has presented ShipFinance.ai, a modular agentic architecture demonstrating how large language models can support loan-origination processes through automated extraction, structured analysis, and standardized application generation. The modular design ensures auditability by separating semantic reasoning (LLM-based extraction from unstructured documents) from numerical computation (predefined financial formulas executed in analysis modules). Preliminary testing demonstrated the technical feasibility of using LLMs to extract key financial metrics
from term sheets and charter agreements, and the integration with the cash flow model. The architecture's value lies not in replacing human expertise but in enabling analysts to focus their time on high-value judgment tasks rather than mechanical data extraction.

However, significant challenges remain before ShipFinance.ai can be used in a production setting. The lender acceptance of standardized, AI-generated applications is uncertain. The regulatory classification of AI-assisted systems under the EU Artificial Intelligence Act and related financial-services regulations remains unsettled as it depends critically on the deployment context. If ShipFinance.ai is used by borrowers to prepare applications, it may function primarily as a data extraction and document-generation tool outside the high-risk category. If integrated into lenders' underwriting processes to evaluate creditworthiness or prioritize applications, it most likely meets high-risk criteria, requiring formal risk-management systems, data quality verification, comprehensive technical documentation, transparency, and human-oversight mechanisms. The latest standards in data-privacy and cybersecurity must be followed as the suggested architecture processes some of the most commercially valuable information that shipping companies possess.

\section{Acknowledgements}
We would like to thank LlamaIndex for accepting our spin-off ShipFinance.ai in their startup program and providing us with the computational resources to verify the concepts presented in this paper. We would also like to thank Olaf Danckwerts for the many inspiring discussions and Alan Albert Piovesana for his valuable feedback.

\appendix

\paperbibliography{refs}

\end{document}

%% file: fig_architecture.tex
\begin{figure}[htbp]
\centering
\begin{tikzpicture}[
  font=\small, >=Stealth, thick,
  extractbox/.style={
    draw, fill=blue!8, rounded corners=5pt,
    minimum width=2.05cm, minimum height=5.0cm,
    text width=2.05cm, align=center
  },
  modbox/.style={
    draw, fill=teal!12, rounded corners=4pt,
    minimum width=2.4cm, minimum height=0.82cm,
    text width=2.2cm, align=center, font=\scriptsize
  },
  cfbox/.style={
    draw, fill=blue!15, rounded corners=4pt,
    minimum width=2.4cm, minimum height=0.52cm,
    text width=2.2cm, align=center, font=\scriptsize
  },
  apibox/.style={
    draw, fill=orange!5, rounded corners=4pt,
    minimum width=10.5cm, minimum height=1.0cm,
    align=center, font=\scriptsize
  },
  chatbox/.style={
    draw, fill=purple!5, rounded corners=4pt,
    minimum width=14.5cm, minimum height=1.0cm,
    align=center, font=\scriptsize
  },
  outbox/.style={
    draw, fill=blue!15, rounded corners=5pt,
    minimum width=2.35cm, minimum height=5.0cm,
    text width=2.1cm, align=center
  },
]


\begin{scope}[on background layer]
  \draw[dashed, rounded corners=6pt, fill=gray!3]
    (-0.65,-2.68) rectangle (3.10,2.88);
  \node[draw=gray!20, fill=gray!7, rounded corners=1pt, thin,
        minimum width=3.0cm, minimum height=4.25cm] at (1.20,-0.18) {};
  \node[draw=gray!33, fill=gray!12, rounded corners=1pt, thin,
        minimum width=3.0cm, minimum height=4.25cm] at (1.10,-0.09) {};
\end{scope}

\node[draw=none, fill=white, rounded corners=1pt,
      minimum width=3.0cm, minimum height=4.25cm,
      inner sep=0pt] (doc) at (1.0,0) {};

\begin{scope}
  \clip (-0.50,-2.125) rectangle (2.50,2.125);

  \fill[gray!65] (-0.20,1.685) rectangle (2.00,1.825);

  \fill[gray!23] (-0.20,1.405) rectangle (1.90,1.505);
  \fill[gray!23] (-0.20,1.235) rectangle (1.50,1.335);
  \fill[gray!23] (-0.20,1.065) rectangle (1.75,1.165);

  \fill[gray!50] (-0.20,0.785) rectangle (1.20,0.915);

  \fill[gray!23] (-0.20,0.535) rectangle (2.00,0.635);
  \fill[gray!23] (-0.20,0.365) rectangle (1.55,0.465);

  \draw[gray!35, thin] (-0.10,-0.125) -- (1.45,-0.125);
  \draw[gray!35, thin] (-0.10,-0.125) -- (-0.10, 0.200);
  \fill[teal!40] (-0.05,-0.125) rectangle (0.22, 0.135);
  \fill[teal!40] ( 0.34,-0.125) rectangle (0.61, 0.200);
  \fill[teal!40] ( 0.73,-0.125) rectangle (1.00, 0.030);
  \fill[teal!40] ( 1.12,-0.125) rectangle (1.39, 0.115);

  \fill[gray!50] (-0.20,-0.475) rectangle (1.10,-0.345);

  \fill[gray!23] (-0.20,-0.725) rectangle (1.90,-0.625);
  \fill[gray!23] (-0.20,-0.895) rectangle (1.55,-0.795);

  \fill[gray!17] (0.00,-1.205) rectangle (1.80,-1.055);
  \draw[gray!30, thin, xstep=0.60cm, ystep=0.15cm]
    (0.00,-1.505) grid (1.80,-1.055);

  \fill[gray!23] (-0.20,-1.745) rectangle (1.80,-1.645);

\end{scope}

\draw[gray!55, rounded corners=1pt] (-0.50,-2.125) rectangle (2.50,2.125); 
\coordinate (idcTop)      at (1.2, 2.125);
\coordinate (indocRibbon) at ($(idcTop)+(0, 0.40)$); 
\node[black, anchor=center, inner sep=0.30cm]
  at ($(idcTop)!0.78!(indocRibbon)$) {\scriptsize \textbf{Input Documents}};

\node[extractbox] (extract) at (5.2,0)
  {\textbf{Value}\\[4pt]\textbf{Extraction}\\[6pt](LLM-based)};

\begin{scope}[shift={(-0.45,0)}]
\node[cfbox]  (cf)      at (9.2,  1.8) {\textbf{Cash Flow}};
\node[modbox] (carbon)  at (9.2,  0.8) {\textbf{Energy Efficiency}\\\textbf{\& Emissions}};
\node[modbox] (charter) at (9.2, -0.1) {\textbf{Revenue}};
\node[modbox] (ratio)   at (9.2, -1.0) {\textbf{Asset}};
\node[modbox] (other)   at (9.2, -1.9) {Further\\Modules};

\coordinate (modtitletop) at ($(cf.north)+(0, 0.40)$);
\coordinate (modenvPadEast) at ($(cf.east)+(0.25,0)$);
\begin{scope}[on background layer]
  \node[draw, dashed, rounded corners=6pt, fill=gray!3, inner sep=0.30cm,
        fit=(cf)(carbon)(charter)(ratio)(other)(modtitletop)(modenvPadEast)] (modenv) {};
\end{scope}
\node[black, anchor=center]
  at ($(cf.north)!0.78!(modtitletop)$) {\scriptsize \textbf{Analysis Modules}};
\end{scope}

\node[apibox] (apis) at (5.5,-3.8)
  {\textbf{External Data Services/Providers}};

\node[outbox] (output) at (12.0,0) {\textbf{Application}\\[3pt]\textbf{Composer}\\(LLM-based)};

\begin{scope}[shift={(14.12,0)}]
  \begin{scope}[on background layer]
    \node[draw=gray!26, fill=gray!8, rounded corners=1pt, thin,
          minimum width=3.0cm, minimum height=4.25cm] at (1.08,-0.12) {};
  \end{scope}
  \node[draw=none, fill=white, rounded corners=1pt,
        minimum width=3.0cm, minimum height=4.25cm,
        inner sep=0pt] (outdoc) at (1.0,0) {};

  \begin{scope}
    \clip (-0.50,-2.125) rectangle (2.50,2.125);

    \fill[gray!65] (-0.20,1.685) rectangle ( 0.90,1.825);
    \fill[teal!22] ( 1.08,1.685) rectangle ( 2.00,1.825);
    \draw[gray!45, thin] (0.99,1.685) -- (0.99,1.825);

    \fill[gray!50] (-0.20,1.392) rectangle (2.00,1.527);

    \fill[gray!23] (-0.20,1.115) rectangle (0.93,1.267);
    \fill[gray!23] (-0.20,0.998) rectangle (0.82,1.068);
    \fill[gray!23] (-0.20,0.820) rectangle (0.93,0.902);

    \fill[gray!17] (1.06,0.820) rectangle (2.00,1.267);
    \fill[teal!40] (1.10,1.185) rectangle (1.96,1.246);
    \fill[teal!40] (1.10,1.035) rectangle (1.72,1.096);
    \fill[teal!38] (1.10,0.862) rectangle (1.94,0.918);

    \fill[gray!14] (-0.20,-0.12) rectangle (2.00,0.488);
    \draw[gray!38, thin, rounded corners=1pt]
      (-0.20,-0.12) rectangle (2.00,0.488);
    \draw[gray!35, thin] (0.90,-0.10) -- (0.90, 0.46);
    \fill[gray!23] (-0.14,0.267) rectangle ( 0.70, 0.402);
    \fill[gray!23] (-0.14,0.025) rectangle ( 0.55, 0.160);
    \fill[gray!23] ( 1.02,0.267) rectangle ( 1.98, 0.402);
    \fill[gray!23] ( 1.02,0.025) rectangle ( 1.76, 0.160);

    \fill[gray!23] (-0.20,-0.34) rectangle (1.90,-0.22);
    \fill[gray!23] (-0.20,-0.52) rectangle (1.88,-0.39);

    \fill[gray!50] (-0.20,-0.83) rectangle (2.00,-0.68);
    \fill[gray!23] (-0.20,-0.98) rectangle (1.92,-0.86);
    \fill[gray!23] (-0.20,-1.18) rectangle (1.76,-1.05);

  \end{scope}

  \draw[gray!55, rounded corners=1pt] (-0.50,-2.125) rectangle (2.50,2.125);

  \coordinate (orcTop) at (1.0, 2.125);
  \coordinate (orcRibbon) at ($(orcTop)+(0, 0.40)$);
  \node[black, anchor=center, inner sep=0.30cm]
    at ($(orcTop)!0.78!(orcRibbon)$) {\scriptsize \textbf{Standardised Application}};
\end{scope}

\draw[->, semithick] (output.east) -- (outdoc.west);

\node[chatbox] (chatbot) at (6.6, 4.2)
  {\textbf{Chatbot Interface} (LLM-based)};

\draw[<-, dashed, thin]
  (1.2, 2.88) -- node[right=2pt, font=\scriptsize] {Upload Documents}
  (1.2, 2.88 |- chatbot.south);

\draw[<->, dashed, semithick]
  (extract.north) -- node[right=2pt, font=\scriptsize] {Context}
  (extract.north |- chatbot.south);

\draw[->, dashed, semithick]
  (modenv.north) -- node[right=2pt, font=\scriptsize] {Report}
  (modenv.north |- chatbot.south);

\draw[<->, dashed, semithick]
  (output.north) -- node[right=2pt, font=\scriptsize] {Report / Complete}
  (output.north |- chatbot.south);
\draw[->, semithick]
  ($(doc.east)+(0, 1.28)$)
  -- node[pos=0.40, anchor=west, yshift=4pt, font=\tiny\itshape, gray!65] {text}
  ($(extract.west)+(0, 1.28)$);
\draw[->, semithick]
  ($(doc.east)+(0, 0.04)$)
  -- node[pos=0.40, anchor=west, yshift=4pt, font=\tiny\itshape, gray!65] {figures}
  ($(extract.west)+(0, 0.04)$);
\draw[->, semithick]
  ($(doc.east)+(0,-1.28)$)
  -- node[pos=0.40, anchor=west, yshift=4pt, font=\tiny\itshape, gray!65] {tables}
  ($(extract.west)+(0,-1.28)$);

\draw[->, semithick] ($(extract.east)+(0,  1.8)$) -- node[pos=0.7, anchor=east, yshift=5pt, font=\tiny\itshape, gray!65] {values} (cf.west);
\draw[->, semithick] ($(extract.east)+(0,  0.8)$) -- node[pos=0.7, anchor=east, yshift=5pt, font=\tiny\itshape, gray!65] {values} (carbon.west);
\draw[->, semithick] ($(extract.east)+(0, -0.1)$) -- node[pos=0.7, anchor=east, yshift=5pt, font=\tiny\itshape, gray!65] {values} (charter.west);
\draw[->, semithick] ($(extract.east)+(0, -1.0)$) -- node[pos=0.7, anchor=east, yshift=5pt, font=\tiny\itshape, gray!65] {values} (ratio.west);
\draw[->, semithick] ($(extract.east)+(0, -1.9)$) -- node[pos=0.7, anchor=east, yshift=5pt, font=\tiny\itshape, gray!65] {values} (other.west);

\draw[->, semithick] (carbon.east)  to[out=60, in=-60] (cf.east);
\draw[->, semithick] (charter.east) to[out=70, in=-60] (cf.east);
\draw[->, semithick] (ratio.east)   to[out=75, in=-60] (cf.east);
\draw[->, semithick] (other.east)   to[out=80, in=-60] (cf.east);

\draw[->, semithick] (cf.east)      -- ($(output.west)+(0,  1.8)$);
\draw[->, semithick] (carbon.east)  -- ($(output.west)+(0,  0.8)$);
\draw[->, semithick] (charter.east) -- ($(output.west)+(0, -0.1)$);
\draw[->, semithick] (ratio.east)   -- ($(output.west)+(0, -1.0)$);
\draw[->, semithick] (other.east)   -- ($(output.west)+(0, -1.9)$);

\draw[<-, dashed, semithick]
  (modenv.south) -- node[right=2pt, font=\scriptsize] {Pull Structured Data from API}
  (modenv.south |- apis.north);

\draw[<-, dashed, thin]
  (1.2,-2.68) -- node[right=2pt, font=\scriptsize] {Pull Documents from API}
  (1.2,-2.68 |- apis.north);

\end{tikzpicture}
\begingroup
\captionsetup{
  format=plain,
  width=0.8\linewidth,
  justification=justified,
  singlelinecheck=false,
}
\caption{Suggested architecture for an AI-augmented system for preparing a loan application in ship finance. A chatbot interface with a document upload function guides the user through the preparation of a loan application and asks to provide information like the IMO number or existing charter agreements. The value extraction module extracts the information required by the analysis modules from the provided documents. This component is LLM-based to enable in-context extraction from unstructured text. Additional information, such as ship construction details, is retrieved from external data providers. The analysis modules compute the quantitative results required for the application. The Energy Efficiency \& Emissions, Revenue, and Asset modules contribute their results to the cash flow analysis. Finally, the application composer aggregates all outputs into a standardized loan application, without generating unsupported information.}
\label{fig:architecture}
\endgroup
\end{figure}
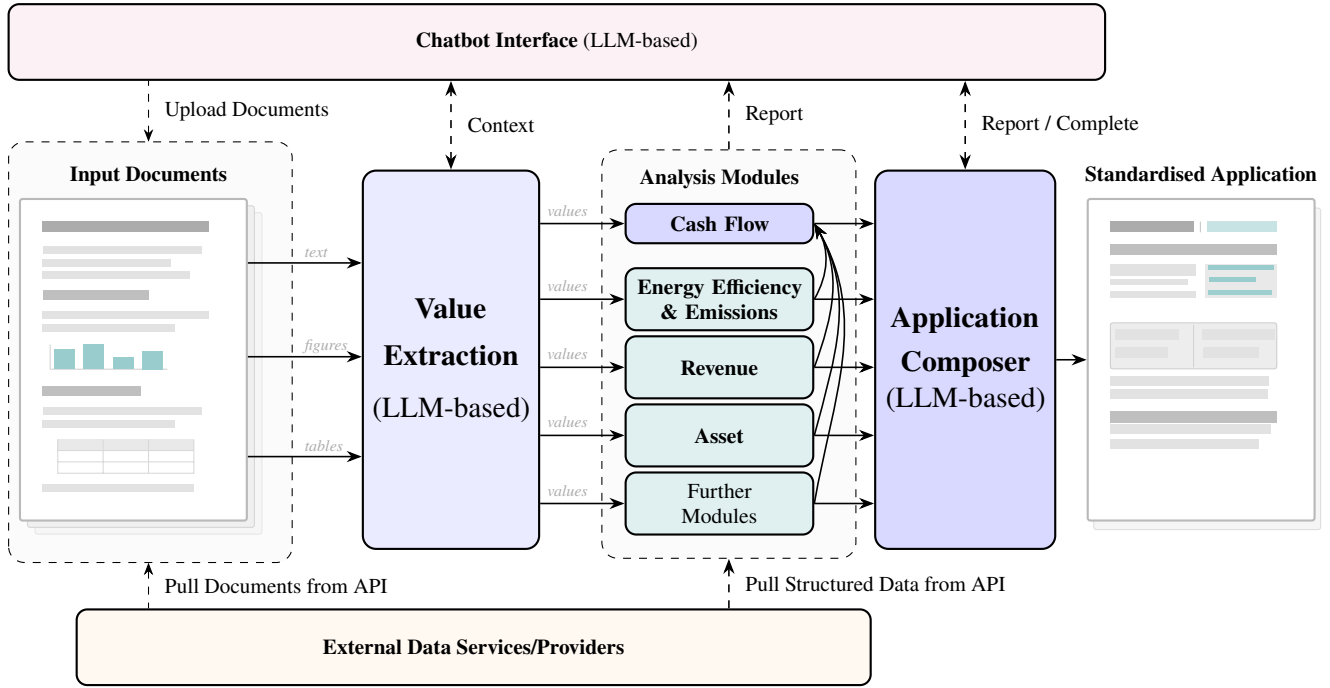